\documentclass[fleqn,twoside]{article}
\usepackage[headings]{espcrc2}

\readRCS
$Id: espcrc2.tex,v 1.2 2004/02/24 11:22:11 spepping Exp $
\usepackage{graphicx}
\usepackage[figuresright]{rotating}

\newcommand{\AmS}{{\protect\the\textfont2
  A\kern-.1667em\lower.5ex\hbox{M}\kern-.125emS}}

\title{CLEO Dalitz plot results}

\author{Jonathan L. Rosner\address{Enrico Fermi Institute and Department of
        Physics, University of Chicago \\ 
        5640 South Ellis Avenue, Chicago, IL 60637 USA}}
       
\runtitle{CLEO Dalitz plot results}
\runauthor{J. Rosner}

\begin{document}

\begin{abstract}
Present and future contributions of the CLEO experiment to the study of $D$
Dalitz plots are presented.  Such Dalitz plots can be of help in determining
weak phases from $B \to D K$ decays.
\vspace{1pc}
\end{abstract}

\maketitle

\section{INTRODUCTION}

$B \to D_{\rm CP} K$ decays are a source of information on the weak phase
$\gamma$.  For $D$ modes such as $K_S \pi^+ \pi^-$, $\pi^+ \pi^- \pi^0$, $K^+
K^- \pi^0$, and $K_S K^\pm \pi^\mp$, Dalitz plots yield information on
CP-eigenstate and flavor-eigenstate modes and their relative phases
\cite{Asner:2003gh}.
This report will briefly review $B \to D_{\rm CP} K$ decays and show
how they determine $\gamma$.  Results from CLEO on the Dalitz plots for the
decays $D^0 \to K^+ K^- \pi^0$, $D^0 \to K_S \pi^+ \pi^-$, and $D^0 \to \pi^+
\pi^- \pi^0$ will be presented, and remaining steps will be described.

\section{THE CLEO-c DETECTOR}

Key elements of the CLEO-c detector \cite{Kubota} are the Ring
Imaging Cherenkov Counter for particle identification, momentum resolution
$\Delta p /p \simeq 0.6\%$ at 1 GeV/$c$ for charged particles, and a CsI
calorimeter with resolution 2.2\% for $E_\gamma = 1$ GeV and 5\% at 100 MeV.
The detector makes use of a new all-stereo inner drift chamber.

\section{$\gamma$ FROM $B \to D K$ DECAYS}

The interference of $b \to c \bar u s$ (real) and $b \to u \bar c s ~(\sim \!
e^{-i \gamma})$ subprocesses in $B^- \to D^0 K^-$ and $B^- \to \overline{D}^0
K^-$, respectively, is sensitive to the weak phase $\gamma$.  Graphs for the
two processes are a tree graph $\sim V_{cb}V^*_{us}$ and a color-suppressed
graph $\sim V_{ub}V^*_{cs} r$ ($r \ll 1$).  This interference may be probed by
studying common decay products of $D^0$ and $\overline{D}^0$ into neutral $D$
CP eigenstates or into doubly-Cabibbo-suppressed modes \cite{Bigi:1988ym}.

\section{$D^0 \to K^+ K^- \pi^0$}

The four rates $B^\pm \to K^\pm (K^{*+} K^-)_D$ and $B^\pm \to
K^\pm (K^{*-} K^+)_D$ provide information on $\gamma$ if the relative (strong)
phase between $D^0 \to K^{*+} K^-$ and $D^0 \to K^{*-} K^+$ is known
\cite{Grossman:2002aq}.  One can learn this relative phase from the study of
$D^0 \to K^+ K^- \pi^0$ since both final states occur and interfere with one
another where $K^{*+}$ and $K^{*-}$ bands cross on the Dalitz
plot \cite{Rosner:2003yk}.

\begin{figure}[t]
\includegraphics[width=0.92 \columnwidth]{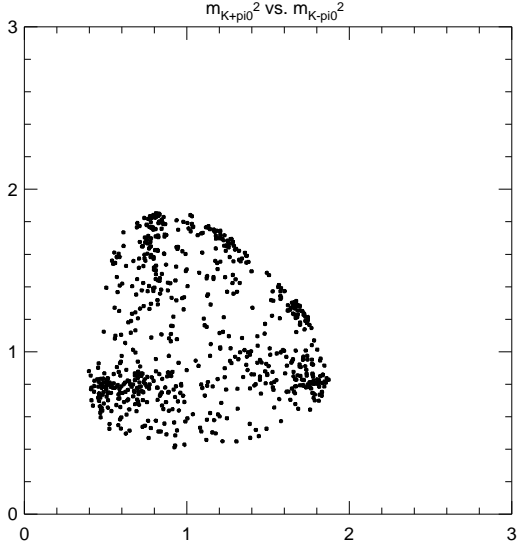}
\vskip -0.3in
\caption{Dalitz plot for $D^0 \to K^+ K^- \pi^0$ \cite{Naik:2005}.  The $D^0$
is tagged via $D^{*+} \to \pi^+ D^0$.  Inclusion of charge-conjugate states is
implied.
\label{fig:paras}}
\vskip -0.2in
\end{figure}

A sample of $D^0 \to K^+ K^- \pi^0$ decays based on 9 fb$^{-1}$
at CLEO III (the predecessor to CLEO-c with an inner silicon vertex
detector instead of the inner drift chamber) shows this interference
clearly \cite{Naik:2005}.  The Dalitz plot is shown in Fig.\ \ref{fig:paras}.
The $K^{*+}$ and $K^{*-}$ bands are found to interfere destructively where they
cross.  The deficit in the $M(K \pi)$ projection above $M[K^*(892)]$ is
difficult to fit using known resonances.  The diagonal band at the upper right
of the plot is the $\phi(1020)$ resonance, with fit fraction (as defined in
\cite{Asner:2003uz}) $\simeq 10\%$.

These data were taken at an $e^+ e^-$ energy at or near the $\Upsilon(4S)$.
The $K^{*+}$ band is $\simeq 3$--4 times as strong as the $K^{*-}$ band,
reflecting differences in form factors and $f_{K^*} > f_K$.  Note the opposite
signs of interference with background on the low sides of the $K^{*+}$ and
$K^{*-}$ bands.  Fits with Breit-Wigner and K-matrix forms are in progress.

\section{$D^0 \to K^0_S \pi^+ \pi^-$}

One can determine $\gamma$ using $B^\pm \to D K^\pm$ followed by (e.g.) $D \to
K_S \pi^+ \pi^-$, $K_S K^+ K^-,~K_S \pi^+ \pi^- \pi^0$ \cite{Giri:2003ty}.  The
method uses interference between $b \to c \bar u s$ and $b \to u \bar c s$
subprocesses.  Its advantages are:  (1) Resonances entail large strong phases,
useful for direct CP asymmetries; (2) one uses only Cabibbo-favored $D$ decay
modes; and (3) one can consider final states involving only charged particles
(e.g., $K_S \pi^+ \pi^-$).  The method has been utilized by Belle 
\cite{Poluektov:2004mf}, yielding $\phi_3 [= \gamma] = (68^{+14}_{-15} \pm 13
\pm 11)^\circ$ for $D K$ and $D^* K$ modes combined based on 275 M $B \bar B$
pairs, and by BaBar \cite{Aubert:2005yj}, yielding $\gamma = (67 \pm 28 \pm 13
\pm 11) ^\circ$ for $D K$, $D^* K$, and $D K^*$ modes combined, based on 227 M
$B \bar B$ pairs.  In each case the last error corresponds to uncertainty
in $D$ decay modeling, which CLEO can help to resolve \cite{Asner:2005}.

\begin{figure}[t]
\includegraphics[width=0.94\columnwidth]{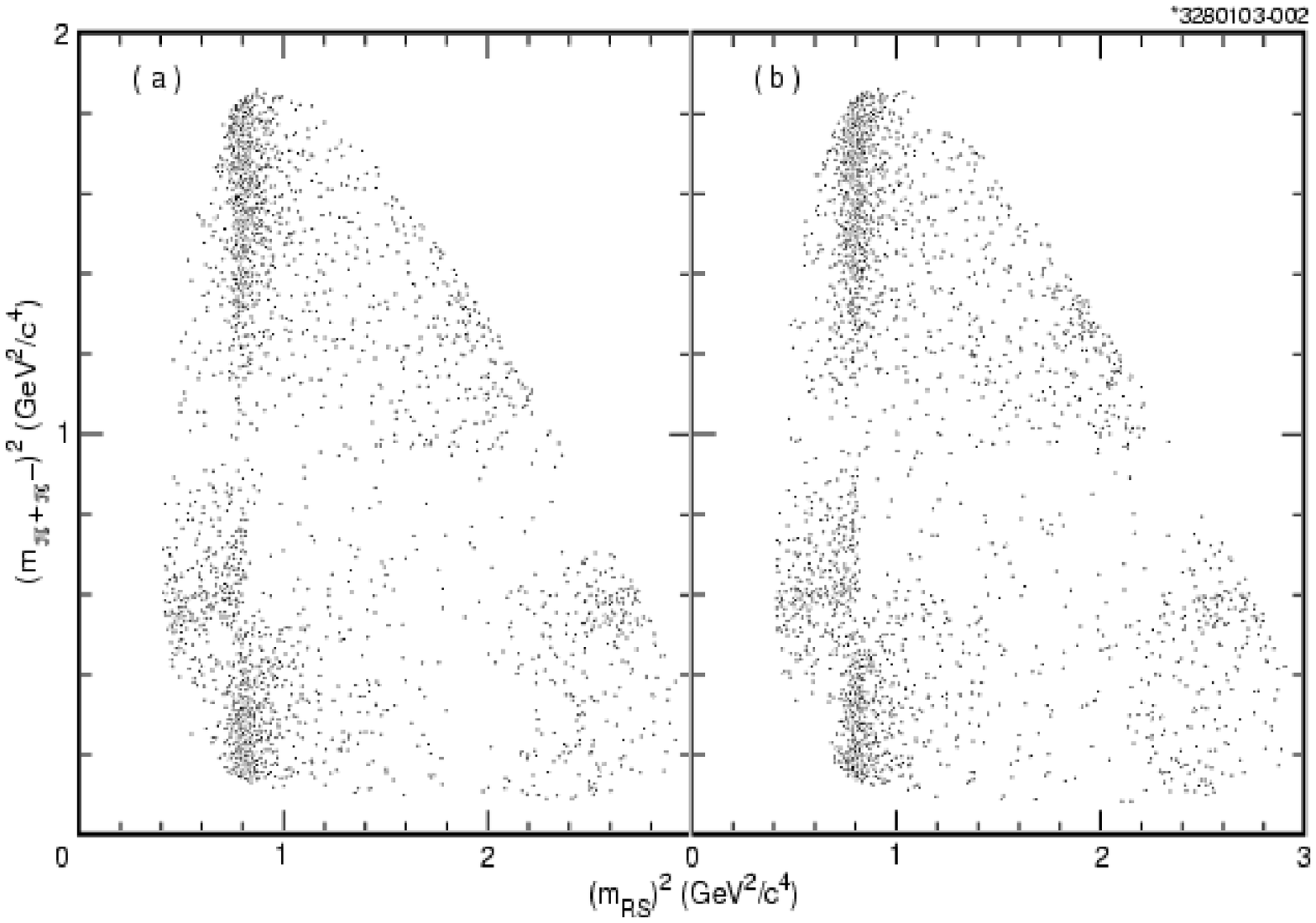}
\vskip -0.3in
\caption{Dalitz plots for (a) $D^0 \to K_S^0 \pi^+ \pi^-$ and
(b) $\overline{D}^0 \to K_S^0 \pi^+ \pi^-$ \cite{Asner:2003uz}.
\label{fig:kppdal}}
\vskip -0.2in
\end{figure}

The Dalitz plots for $D^0 \to K_S^0 \pi^+ \pi^-$ and $\overline{D}^0 \to K_S^0
\pi^+ \pi^-$ obtained from 9.0 fb$^{-1}$ at the CLEO II.V detector are shown
in Fig.\ \ref{fig:kppdal} \cite{Asner:2003uz}.
The doubly-Cabibbo-suppressed $D^0 \to K^{*+} \pi^-$ mode is visible through
destructive interference with the Cabibbo-favored $D^0 \to K^{*-} \pi^+$
``right-sign'' mode.  The dominant fit fractions are $K^{*-}(892) \pi^+$
($\sim 2/3$) and $\overline{K}^0 \rho^0$ ($\sim 27\%$).   The latter is CP-odd
for $\overline{K}^0$ detected as $K_S^0$.  CP-even modes such as $K_S^0
f_2(1270)$, $K_S^0 f_0(980,1370)$ give rise to a fit fraction of $\sim 1/3$.
As usual in such analyses, fit fractions need not add up to 100\%.

\begin{figure}[t]
\includegraphics[width=0.92\columnwidth]{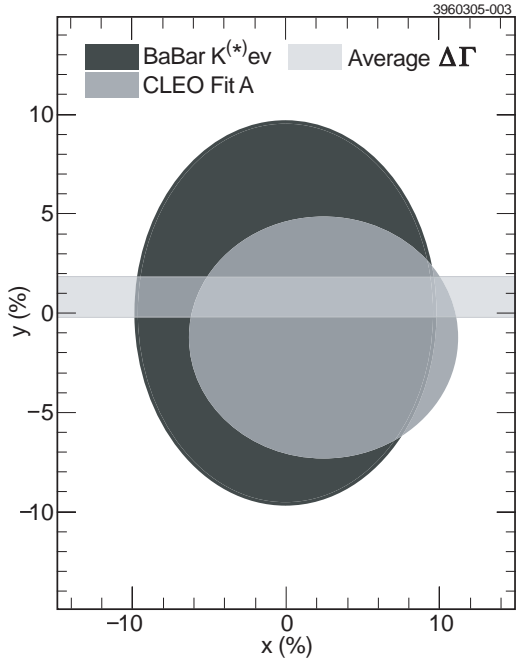}
\vskip -0.3in
\caption{Constraints in the $x$--$y$ plane from various experiments on
$D^0$--$\overline{D}^0$ mixing.
\label{fig:moon}}
\vskip -0.2in
\end{figure}

One can use the Dalitz plot of $D^0 \to K^0_S \pi^+ \pi^-$ to search for
$D^0$--$\overline{D}^0$ mixing \cite{Asner:2005sz}.  We define $\Gamma \equiv
\frac{\Gamma_1+\Gamma_2}{2}$, $x \equiv \frac{m_1 - m_2}{\Gamma}$, $y  \equiv
\frac{\Gamma_1 - \Gamma_2}{2}$.  Mass eigenstates evolve in time as
$|D_{1,2}(t) \rangle= |D_{1,2}(0) \rangle e^{[-i(m_{1,2} - \frac{i\Gamma_{1,2}}
{2})t]}$.  One tags $D^0(t=0)$ flavor using $D^{*+} \to \pi^+ D^0$.  A
time-dependent Dalitz plot fit then is sensitive to $D^0$--$\overline{D}^0$
mixing, leading to the CLEO 95\% c.l.\ moon-shaped region shown in Fig.\
\ref{fig:moon}.  In the fit shown (``A'') no CP conservation was assumed.

\section{$D^0 \to \pi^+ \pi^- \pi^0$}

\begin{figure}[t]
\includegraphics[width=0.98\columnwidth]{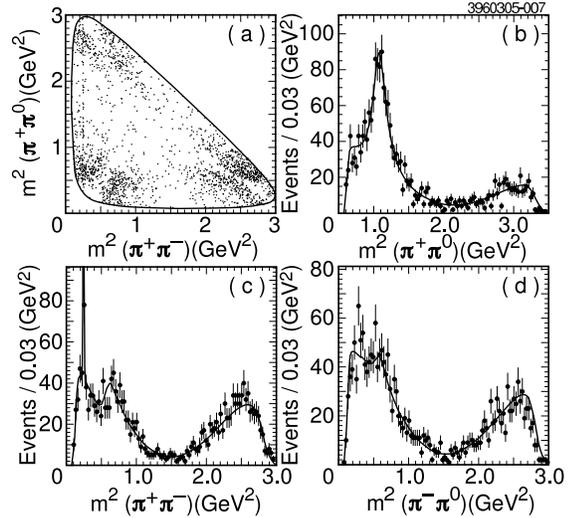}
\vskip -0.3in
\caption{(a) $D^0 \to \pi^+ \pi^- \pi^0$ Dalitz plot; (b) $\pi^+ \pi^0$
projection; (c) $\pi^+\pi^-$ projection; (d) $\pi^-\pi^0$ projection.
\label{fig:dal3pi}}
\vskip -0.2in
\end{figure}

A search for CP violation and a study of the $\pi \pi$ S-wave in the decay $D^0
\to \pi^+ \pi^- \pi^0$ has been performed using 9.0 fb$^{-1}$ of CLEO II.V data
\cite{Cronin-Hennessy:2005sy}.  The corresponding Dalitz plot and some of its
projections are shown in Fig.\ \ref{fig:dal3pi}.  In the projections, points
correspond to data while curves are based on a fit describing the S waves
with a K-matrix following the formalism of \cite{Au:1986vs}.

The CP asymmetry as defined in Ref.\ \cite{Asner:2003uz} is found to be
${\cal A}_{\rm CP} = 0.01^{+0.09}_{-0.07}\pm0.05$.  This mode may be useful
in analyzing $B$ decays since $B^- \to D_{\pi^+ \pi^- \pi^0} K^-$ has been
seen \cite{Aubert:2005hi}, with a sample of 133 events in 229 M $B \bar B$
pairs, corresponding to a combined branching ratio of ${\cal B} = (5.5 \pm 1.0
\pm 0.7) \times 10^{-6}$.  The decay asymmetry in this mode has been found to
be $A = 0.02 \pm 0.16 \pm 0.03$.

\begin{figure}[t]
\includegraphics[width=0.92\columnwidth]{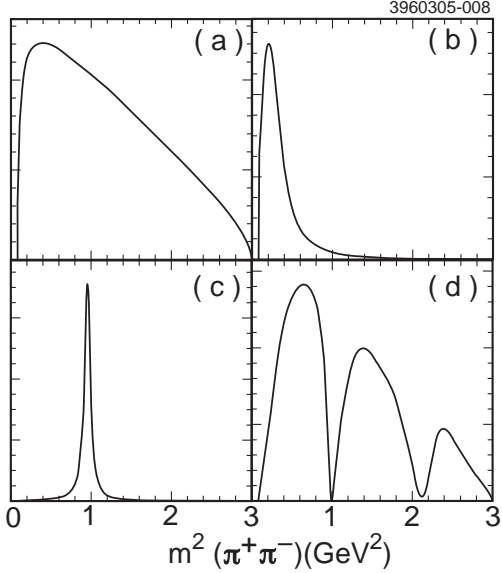}
\vskip -0.4in
\caption{Fits to S-wave component in $D^0 \to \pi^+ \pi^- \pi^0$.  (a)
Flat non-resonant; (b) Breit-Wigner $\sigma(500)$; (c) Breit-Wigner $f_0(980)$;
(d) K-matrix.
\label{fig:3pifits}}
\vskip -0.2in
\end{figure}

All fits to the $D^0 \to \pi^+ \pi^- \pi^0$ Dalitz plot are dominated by the
$\rho^\pm$ and $\rho^0$ bands, with negligible S-wave contributions.  These are
found to be small and fitted adequately by a variety of parametrizations, as
shown in Fig.\ \ref{fig:3pifits}.  This is in contrast to a large S-wave
contribution found by the FOCUS Collaboration in $D^+ \to \pi^+ \pi^+ \pi^-$
\cite{Link:2003gb}.  This could be due to dominance of different processes in
the two decays.  The dominant tree contribution to $D^0 \to \pi^+ \pi^- \pi^0$
involves $D^0 \to \rho^+ \pi^-$ with a fit fraction of 75--80\%, while the
dominant tree contribution to $D^+ \to \pi^+ \pi^+ \pi^-$ involves $D^+ \to
\pi^+ d \bar d$, where $d \bar d$ can fragment into an S-wave.  The
corresponding $D^0 \to \pi^0 u \bar u$ or $\pi^0 d \bar d$ decay is
color-suppressed.

\begin{figure}[t]
\includegraphics[width=0.94\columnwidth]{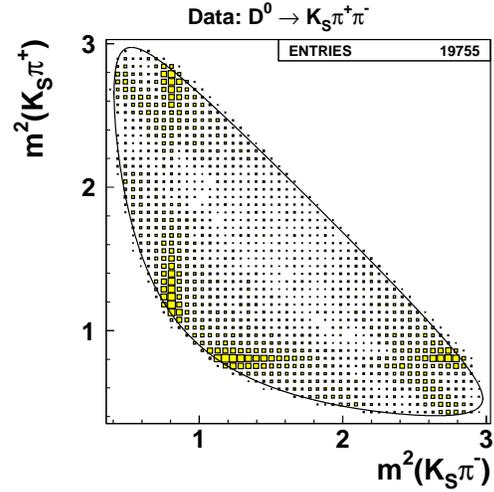}
\vskip -0.4in
\caption{Dalitz plot for $D \to K_S \pi^+ \pi^-$ based on 281 pb$^{-1}$
of CLEO-c data.
\label{fig:kspipi}}
\vskip -0.2in
\end{figure}

\section{REMAINING STEPS}

So far CESR has accumulated $281$ pb$^{-1}$ at the $\psi''(3770)$.  More data
at this energy will provide a clean sample of $D^0 \bar D^0$ and $D^+ D^-$,
with tagging on one side giving the flavor or CP eigenvalue
on the other.  Examples of CP-odd three-body $D^0$ modes are
$K_S \rho$, $K_S \omega$, $K_S \phi$, while some CP-even three-body $D^0$
modes are $K_S f_0(980)$, $K_S f_2(1270)$, $K_S f_0(1370)$.

Once a sufficient sample of $\psi''(3770)$ has been accumulated, it is planned
to run at an energy optimized for $D_s$ production.  Optimization studies are
to begin in August 2005.  Subsequently a year's worth of $J/\psi$ data will
shed light on light-quark and glue states.

In future Dalitz plot analyses the approximate relative phase of $\pi$ between
$K^{*+} K^-$ and $K^{*-} K^+$ in $D^0 \to K^+ K^- \pi^0$ may be used to obtain
$\gamma$ from $B^- \to D^0 K^-$ as suggested in Ref.\ \cite{Grossman:2002aq}.
A sample of $D \to K_S \pi^+ \pi^-$ based 281 pb$^{-1}$ from CLEO-c, for which
the Dalitz plot is shown in Fig.\ \ref{fig:kspipi}, is currently under analysis.
This sample will permit reduction of the Dalitz plot modeling error in
obtaining $\gamma$ from $B^\pm \to D K^\pm$ followed by $D^0 \to K_S \pi^+
\pi^-$ from its current value of $\pm 10^\circ$ to about $\pm 7^\circ$.  One
expects about 100 each of even and odd CP tags in the present sample, and
eventually up to 10 times that if CLEO-c's luminosity goals are met.

The CLEO Collaboration gratefully acknowledges the effort of the CESR staff
in providing us with excellent luminosity and running conditions.
This work was supported by the National Science Foundation
and the United States Department of Energy.  I thank M. Tigner
for hospitality of the Laboratory for Elementary-Particle
Physics at Cornell and the John Simon Guggenheim
Foundation for partial support.

\end{document}